\documentstyle[12pt]{article}
\newcommand\beq {\begin{equation}}
\newcommand\eeq {\end{equation}}
\newcommand\bqa {\begin{eqnarray}}
\newcommand\eqa {\end{eqnarray}}
\newcommand\pr {\partial}
\newcommand\apr {\overline {\partial }}
\newcommand\nn {\nonumber}
\newcommand \z {{\overline z}}
\newcommand\G {GWZW }
\newcommand \NA {\bigtriangledown}
\newcommand \noi {\noindent}
\begin{document}
\bibliographystyle{nphys}

\begin{flushright}
ITEP-93 \\
UUITP,16/1993 \\
hep-th/9305090
\end{flushright}
\vskip 1cm
\centerline{\large\bf Localization in GWZW and Verlinde formula}
\vskip 4 cm

\centerline{\bf A.Gerasimov$^\dagger$}
\vskip 5mm
\centerline{Institute for Theoretical Physics }
\centerline{Thunsbergsvagen 3 Box 803 }
\centerline{ S-75108 Uppsala Sweden }
\vskip 15 mm\noi
\centerline{\bf ABSTRACT}
\vskip 5mm

Gauged Wess-Zumino-Witten theory for compact  groups is considered.
It is shown that this theory has fermionic BRST-like symmetry and may
be exactly solved  using localization approach. As an example we
calculate functional integral for the case of $SU(2)$ group on the
arbitrary Riemann surface. The answer is the particular case of
Verlinde formula for the number of conformal blocks.

\vfill\noi
$^\dagger$
permanent address:ITEP,B.Cheryomushkinskaya 25,Moscow,117259,Russia

\eject

\section{Introduction}

One of the interesting problem in Quantum Field Theory is to
understand  in unifying terms the nature of integrability of various
kinds of known quantum exactly soluble theories. It seems that
quantum exactly soluble theories are in the class of the theories  in
which after possible addition of auxiliary fields  we may use for all
but finite number of modes quasiclassical approximation to get exact
answers. Thus in these theories it is possible to reduce explicitly
all  computations to finite dimensional integral (or even to the
number of one-dimensional integrals) and to the problem to find
solutions of the classical equations of motion. This is apparently
implies some kind of fixed point formula. It is quite desirable to
find the universal method to solve these theories and probably the
best candidate is cohomological approach developed in \cite{At,DH}
and recently reconsidered in  \cite{W3,Niem1,Niem2,Niem3}.

We know several classes of two-dimensional theories to which
localization or cohomological approach is (or probably may be )
applicable. The most natural class consists of topological  theories
with explicit BRST symmetry and the correlation functions of these
theories may be described in terms of topological characteristics of
some finite dimensional geometrical objects.

 We have also trivial class of exactly soluble theories . It is free
theories with quadratic action where quasiclassical approximation is
exact. Conformal field theories give a slightly more complex example
of exactly soluble theories closely connected with free massless
theories. The difference  is in additional  constraints on finite
number of degrees of freedom. This suggests that correlation
functions of any conformal theory on any Riemann surface may be
calculated directly by some version of localization methods. When
coupled to gravity these theories are deeply related with topological
theories and surely admit localization.

It is interesting to consider general 2d integrable theories from
this point of view. Taking into account possible auxiliary fields
these theories may be treated as deformation of conformal theories
which somehow preserves "free field" description but now these fields
are massive (see e.x.\cite{G}). Here  also  different nontrivial
classical solutions are possible and the whole picture is more
complex but the existence of good action-angle variables encouraged
to try to apply localization approach. Surely it is natural to begin
with finite-dimensional integrable systems using the Lagrangian
description \cite{G} and the explicit quantization \cite{OP}

Notice that there is also general construction of BRST-like symmetry
in Hamiltonian formalism for Hamiltonian with rather mild
restrictions which formally gives the credit to quasiclassical
approximation to be exact \cite{Niem1,Niem2,Niem3} but exact
conditions of applicability of this formalism are unclear.

Nice example of application of localization method is given by 2d
Yang-Mills (YM) theory (\cite{W1,W2}). Here it is possible to
reproduce the known answers for correlation functions \cite{Mig} and
to connect YM theory with topological theory which calculates volume
of moduli space of flat connections.

Bellow we will consider slightly more complex example of
Gauged-Wess-Zumino-Witten (\G )\cite{Gav}. It is topological
two-dimensional theory closely related with conformal
Wess-Zumino-Witten model (WZW). In particular its functional integral
gives the number of conformal blocks in WZW. Elegant formula due to
Verlinde \cite{Ver} is known for the number of conformal blocks. We
will show that the functional integral may be exactly calculated and
reproduces the Verlinde formula. As an example we will consider the
case of $SU(2)$. It is interesting to remark that in this calculation
we should treat finite number (one in the case of $SU(2)$) degrees of
freedom  exactly and the rest part quasiclassicaly.

The content is the following . In the second section we describe
simple derivation of the action of \G model with the emphasis on the
connection with 3d Chern-Simons theory \cite{W4}. Section 3 is
devoted to reconsideration of 2d Yang-Mills theory where we pay
attention on the one-loop corrections to classical answer. In section
4 we describe the calculation of functional integral for \G with
group $SU(2)$ and obtain Verlinde formula for the number of conformal
blocks.

\vskip 1cm

\section{Gauged Wess - Zumino Theory (GWZW)}
In this section we will give a short derivation of functional
integral representation for \G theory starting with Chern-Simons (CS)
theory. This is revealed the connection of correlation functions of
\G and correlation functions of CS theory which in turn are connected
with the modular geometry of conformal blocks of WZW. In particular
it will be obvious that vacuum correlation functions of \G are equal
to the dimension of the space of conformal blocks in
Wess-Zumino-Witten model.

Consider CS theory  with compact gauge group $G$ on the three
dimensional  space $M$ with the topology of the product of the
one-dimensional circle and two-dimensional Riemann surface $M=S^1
\times \Sigma $. Let us given the connection $A$ with the curvature
$F$ and some integer $k$.Then using explicit (2+1) notations
functional integral for CS-theory may be represented in the form:

\bqa
Z = \int {\cal D}A\, e^{k\int (A_i\pr _t A_j \epsilon ^{ij} + \int
A_0 F_{i,j}(A) \epsilon ^{ij}) dtd^2 x }
\eqa
where 0-component is along $S^1$ and $x^i$ are coordinates on
$\Sigma$.
It is known (\cite{W4}) that Hilbert space of CS theory is given by
the cohomology groups of the moduli space ${\cal M}^G $ of flat
connections with gauge group $G$ with coefficients in the $k^{th}$
power of the determinant bundle.
\bqa
{\cal H}_{CS} = H^0 ({\cal M}^G ;{\cal L}^{\otimes k})
\eqa
(under the condition the higher cohomology groups are zero).
 On the other side the elements of this cohomology groups  are
corresponded to   conformal blocks in the WZW theory with the
Kac-Moody group ${\widehat G}_k$. CS theory has zero Hamiltonian and
the partition function of CS-theory on $M$ may be interpreted as the
trace of unit operator over the space of conformal blocks and  hence
equal to the dimension of the space :
\bqa
Z = dim {\cal H}_{CS} = dim H^0 ({\cal M}^G ;{\cal L}^{\otimes k})
\eqa

Usually the higher cohomology are zero and the dimension of the zero
cohomology group may be calculated through Riemann-Roch theorem:
\bqa
\sum _p (-1)^p dim H^p ( {\cal M}^G ;{\cal L}^{\otimes k})  = \int
_{{\cal M}^G } Ch({\cal L}) Td(T{\cal M}^G )
\eqa
where Chern character and Todd class in the case of direct sum of
linear bundles $E = \oplus x_i $ are :
\bqa
Ch(E) = \sum_i e^{x_i } \\
Td(E) = \prod _i \frac {x_i }{1-e^{-x_i }}
\eqa
It is obvious that the role of $S^1 $ is rather trivial in the above
considerations and thus we may suppose that the same answer is
obtained in the 2d theory which appears when the radius of the $S^1 $
goes to zero . Bellow we will show that along this line we will
obtain \G theory .

To get the partition function on the $M=S^1 \times \Sigma $ we will
consider the propagator $K$ in CS theory. From the structure of the
action of CS theory we could easily derive that the propagator may be
represented as a product of the propagator in the theory defined by
the functional integral:
\bqa
Z_{FreeCS} = \int {\cal D}A e^{k\int A_i\pr _t A_j \epsilon ^{i,j}}
\eqa
and the projector on the states with zero two-dimensional curvature.
To construct both peaces let us define the scalar product on the
space of wave functions over space of all connections :
\bqa
<\Psi _1 |\Psi _2 > = \int {\cal D}A_z  {\cal D}A_{{\overline
z}}e^{(-k\int A_z A_{{\overline z}} )} \Psi \left ( A_z \right )
{\overline {\Psi (A_{{\overline z}})}}
\eqa
where $\Psi \left ( A_z \right ) $ is wave function. For such scalar
product propagator in the theory without constraints is :
\bqa
K_{Free CS}(A^1 _z ,A^2 _{\z}) = e^{-k\int A^1 _z A^2 _{{\overline
z}} }
\eqa
For example we may verify that

\bqa
K_{Free CS}^2 = K_{Free CS}
\eqa
as it should be for the theory with zero Hamiltonian.

Constraint $F(A_z ,A_{{\overline z}}) = 0 $ for the polarization
when wave function depends only on $A_z $ is:
\bqa
\left (\pr \frac {\delta }{\delta A} - {\overline {\pr}} A + \left [
A , \frac {\delta }{\delta A} \right ] \right ) \Psi (A) = 0 \\ \nn
\eqa
This condition may be interpreted as infinitesimal variant of the
following condition of  "gauge invariance" of wave functions of CS
-theory:
\bqa
&& \Psi \left ( A^g \right ) = e^{k S_{WZW}(g) + k\int
Ag^{-1}{\overline {\pr}g} } \Psi \left ( A \right ) \\
&& A^g = g^{-1}dg +g^{-1}Ag
\eqa \label{inv}
\noi
where $S_{WZW}$ is the standard action of WZW theory. Using
Polyakov-Wiegman formula it is  simple to verify that the following
projector operator is compatible with the condition (\ref{inv}).
\bqa
\Pi \Psi (A) = \int {\cal D}g  e^{k S_{WZW}(g) + k\int
Ag^{-1}{\overline {\pr}} g} \Psi \left ( A^g \right )
\eqa
We normalize the projector to have the necessary property for any
projector:
\bqa
\Pi ^2 = \Pi
\eqa
Now we may calculate  the partition function of CS theory :
\bqa
&& Z = Tr K_{Free CS} \Pi = \int {\cal D}A_z  {\cal D}A_{{\overline
z}}{\cal D}g e^{-k\int A_z A_{{\overline z}}+ k S_{WZW}(g) } \nn \\
&& e^{ \int Ag^{-1}{\overline {\pr}g} + k\int A_z A_{{\overline z}}
^g } = \int {\cal D}A_z  {\cal D}A_{{\overline z}}{\cal D}g e^{k
S_{GWZW}(g,A_z ,A_{{\overline z}} ) } \label{fGWZW}
\eqa
This is the functional integral for Gauged Wess-Zumino-Witten theory
with the action:
\bqa
&& S_{\G}(g,A) =kS_{WZW}(g) + \\
&& \frac {i}{2\pi} k(\int Ag^{-1}{\overline {\pr}g} + g\pr g^{-1}
A_{\z} +gA_z g^{-1}A_{\z} - A_z A_{\z}) \nn \\ \nn
\eqa
Let us mention another interesting representation for the \G as  the
ratio of two determinants. Taking into account the expression for
regularized determinant:
\bqa
det \Delta \left [ A\right ] = e^{S_{WZW}\left [ A\right ]
+S_{WZW}\left [ {\overline A}\right ] + \int A{\overline A}}
\eqa
we have the following representation:
\bqa
e^{S_{\G}(g,A_z ,A_{\z})} = \frac {det \Delta \left [ A,{\overline
A^g }\right ] }{det \Delta \left [ A,{\overline A}\right ] } \\ \nn
\\
{\overline A^g } = g^{-1}\apr g + g^{-1}{\overline A}g
\eqa
\vskip 5 mm
To summarize we have 2d theory with the property that its functional
integral gives the number of conformal blocks in WZW theory. It
appears that for this quantity there exist rather simple
representation. Using the nontrivial fact that structure constants of
fusion algebra $N_{ij}^k $:
\bqa
\phi _i \times \phi _j = \sum _k N_{ij}^k \phi _k
\eqa
are diagonalized by the matrix of modular transformation of one loop
conformal blocks Verlinde (\cite{Ver}) deduce formula for the number
of conformal blocks of arbitrary conformal theory on the surface
$\Sigma $ of genus g through the matrix of one-loop modular
transformation $S_{ij}$:
\bqa
dim V_g = tr \left ( \sum _{i=0} ( N^i )^2 \right ) ^{g-1} = tr \left
( \sum _{i=0} \frac {(S_i ^h )^2 }{(S_0 ^h  )^2 }\right ) = \sum
_{n=0}^{N-1} |S_{n0}|^{2(1-g)}
\eqa

Im particular case of WZW with the group ${\widehat G}_k$ it has the
form:
\bqa
dimV_{\Sigma }^{g} = (C(k+h)^r )^{g-1} \sum _{\lambda \in P^k _{+}}
\prod _{\alpha \in \Delta } \left (1-e^{2\pi i\alpha \left (\frac {
\lambda +\rho }{k+h} \right )}\right )^{1-g}
\eqa
where $C$ is the order of the center and $h$ - Coxeter number, $\rho
$ is the half of the sum of positive roots and $P_{+} ^k $ - weights
of integrable representations of ${\widehat G_k }$. Below we derive
this formula for the case of ${\widehat SU(2)}_k $ on arbitrary
surface. In this case the number of conformal blocks is given by
\bqa
dimV^g _{\Sigma} = \sum _{n=0}^{N-1} |S_{n0}|^{2(1-g)} = \sum _{n=0}
^{k+1} \left (\frac {k+2}{2}\right ) ^{g-1} \left ( sin \left (\frac
{n+1}{k+2}\pi \right ) \right )^{2(1-g)}
\eqa
\vskip 1cm
Bellow we will show how this formula may be derived from functional
integral (\ref{fGWZW})
\vskip 1 cm
\section{\bf Two-dimensional Yang-Mills theory}

In this section we reconsider the calculation of functional integral
for 2d Yang-Mills theory described in \cite{W1,W2}. The difference
with \cite {W1,W2} will be in taking into account non-trivial
one-loop contribution which provide us with the exact result.

Let $A$ be the connection on the surface $\Sigma$ with the curvature
$F=dA + \frac {1}{2} \left [A,A \right ] $  and $\phi$ is the scalar
field in adjoint representation. Then the action of two-dimensional
Yang-Mills theory may be written as:
\bqa
S = tr \int (\phi F + \frac {1}{2}\psi \wedge \psi + \epsilon \phi ^2
)
\eqa
where we introduce the fermions $\psi$ to have a canonical measure
$dAd\psi $

This theory  posses  the following BRST-like symmetry:
\bqa
&& \delta A _{\mu } = \lambda \psi _{\mu } \\
&& \delta \psi  _{\mu } = \lambda D_{\mu } \phi = \\
&& \lambda (\pr _{\mu } \phi + \left [ A_{\mu },\phi ]\right ]) \\
&& \delta \phi = 0
\eqa \label{BR1}
We may define an invariant observables by the condition that if the
observable is represented as the integral of some operator-valued
form over closed cycle the BRST variation of this form should be
exact \cite{W3}:
\bqa
&& \left [Q,{\cal O}^{(0)}_n \right ] = 0  \\
&& \left [Q,{\cal O}^{(1)}_n \right ] = d{\cal O}^{(0)}_n   \\
&& \left [Q,{\cal O}^{(2)}_n \right ] = d{\cal O}^{(1)}_n   \\
\label{des}
\eqa

{}From (\ref{BR1}) we have obvious zero-form observables ${\cal
O}^{(0)}_n = tr \phi ^n $ and from (\ref{des}) we find the whole
family of observables:
\bqa
{\cal O}^{(0)}_n = tr \phi ^n  \\
{\cal O}^{(1)}_n = ntr \phi ^{n-1} \psi \\
{\cal O}^{(2)}_n = ntr\phi ^{n-1} F + \frac {n}{2} tr \sum
^{n-2}_{k=0} \phi ^k \psi \wedge \phi ^{n-k-2} \psi \\
\eqa

To calculate the functional integral which respect some fermionic
symmetry we should find such  fermionic function $V$ that  deformed
action:
\bqa
S = S_0 + \beta \left [ Q,V \right ]
\eqa
with $\beta \rightarrow \infty $ gives rise to concentration of the
functional integral on some  small finite dimensional subspace of the
field configurations. The useful choice in our situation is;
\bqa
&& \delta S = \left [ Q,V \right ] \\
&& V = \beta \int \Psi (\NA \phi -\NA F ) \\
&& \delta S = \beta \int \NA \phi (\NA \phi - \NA F ) + \phi \Psi
\wedge \Psi  + \Psi \Delta \Psi +F \Psi \wedge \Psi = \\
&& \beta \int (\NA \phi - \NA F )^2 - |\NA F|^2 + (\phi - F)\Psi
\wedge \Psi + \Psi \Delta \Psi
\eqa \label{GF1}
Let us note  that by very general arguments \cite{W5} functional
integral with fermionic symmetry concentrates in infinitesimal
neighborhood of the zero locus of symmetry transformations. Thus we
may suppose  from explicit form (\ref{BR1}) that quasiclassical
approximation is exact for all modes with exception of zero modes of
the field $\phi$ which should be treated exactly. This is in
agreement with the (\ref{GF1}). It is useful to  divide  the field
$\phi $ on constant and nonconstant parts:
\bqa
\phi =\phi_0 + \phi _{\ast}
\eqa
Then bosonic part of the action has the form:
\bqa
\int (\phi F + \epsilon \phi ^2 )= \int ((\phi_0 + \phi _{\ast}) F +
\epsilon \phi _{0}^2 +\epsilon \phi _{\ast}^2 )
\eqa
Classical equations of motions are :
\bqa
\Psi = 0  \nn \\
\NA \phi _{\ast} =0  \\
F_{\ast} + 2\epsilon \phi _{\ast} = 0  \nn
\eqa
As the consequence we have:
\bqa
\NA F =0
\eqa
With the appropriate gauge transformation  in $SU(2)$ case we may
choose representative which has only one non-zero component of $\phi
$ along $\sigma _3 $ direction (for example). From the condition on
$\phi $ to be covariantly constant we deduce that $\phi $ is constant
($\phi _{\ast} =0$) and ($\pm $)- components (with respect to
decomposition of $sl(2)$ on lower-triangle (+) ,diagonal and
upper-triangle (-) parts)  of gauge fields are zero. Hence we are
reduced to  $U(1)$-bundle on the surface and from topological
restrictions the values of the curvature (which is constant) is $2\pi
m $ where $m$ should be integer.

Now consider the second variation of the action.
\bqa
&& \delta ^2 S = tr \int 2\phi ^0 \left [\delta A_{\mu } , \delta
A_{\nu}\right ] +2\delta \phi ^{+}(\NA _{\mu}\delta A_{\nu})^{-}
+2\delta \phi ^{-}(\NA _{\mu}\delta A_{\nu})^{+} + \\
&&  \delta \phi ^{0} \pr _{\mu} \delta A_{\nu} ^0 + 2\epsilon \delta
\phi ^{0} \phi ^{0} +2\epsilon \delta \phi ^{+} \delta \phi ^{-}
+2\epsilon \delta \phi ^{-} \phi ^{+} = \\
&& = tr \int \phi ^{0} \left ( \delta A^{+}_{\mu} - \NA _{\mu} \left
(+ \frac {\delta \phi ^{+}}{2\phi ^{0}}\right )\right )  \left (
\delta A^{-}_{\mu} - \NA _{\mu} \left (- \frac {\delta \phi
^{-}}{2\phi ^{0}}\right )\right ) + \\
&& + \delta \phi ^{0} \pr _{\mu} \delta A_{\nu} 2\epsilon \delta \phi
^{0} \phi ^{0}
\eqa
By shifting integration over connections we get the following
representation for the action up to second order around the classical
solution parametrized by integer $m= \frac {1}{2\pi } \int F $ and
constant mode of the field $\phi $ :
\bqa
S=\int \phi _0 ^{0} 2\pi i m + \int \epsilon (\phi _0 ^0 )^2 + \int
\phi _{0} \delta A^{+}_{\mu} \delta A^{-}_{\nu} + \epsilon (\delta
\phi  ^{0}_{\ast}) ^2 + \delta \phi ^{0}_{\ast} \pr _{\mu} \delta
A_{\nu}^0
\eqa
Taking into account that the action is invariant under the gauge
symmetry we may parametrize $\phi ^{\pm}$ in terms of infinitesimal
gauge parameter$\lambda $:
\bqa
\delta \phi ^{\pm} = \left [\lambda ^{\pm},\phi ^{0} \right ]
\eqa
The space of the classical solutions is reduced  in essence to the
moduli space of $U(1)$ connections and the measure in the vicinity of
the classical solution may be represented in the form:
\bqa
d\phi ^{+}d\phi ^{-}d\phi ^{0}dA^{+}dA^{-}dA^{0}d\Psi ^{+}d\Psi
^{-}d\Psi ^{0} = \\
d\lambda ^{+} d\lambda ^{-} d\lambda ^{0} \left [\frac
{dA^{+}dA^{-}dA^{0}}{d\lambda ^{0}}\right ] d\mu _{J} det \left
[2\phi ^{0} \mbox{{\bf I}}\right ]^2  \\ \nn
\eqa
In the previous expression  $d\mu _{J} $ is the natural measure on
the Jacobian of the curve and {\bf I} is the unit operator.
Taking Gauss integral we get the following "quasiclassical" answer in
the topological sector parametrized by integer $m$:
\bqa
Z_{m} = e^{S_{class}}Vol_{\mbox{Jacobian}} Vol_{\mbox{Gauge group}}
\frac {det \left [2\phi ^{0} {\bf I}\right ]^{2(g-1)}_{(0)}}{det
\left [2\phi ^{0} {\bf I}\right ]^{2(g-1)}_{(1)}}
\eqa
where $det_{0,1}$ are determinants over the space of $0(1)$
differentials. As usual we should divide on the volume of the gauge
group $Vol_{\mbox{Gauge group}}$.
Let us compare accurately the number of eigenvalues of the
1-differential and 0-differential eigenvalues.
Consider the differential operator
\bqa
{\overline D}={\overline \pr }+{\overline A }:\Omega ^0 \rightarrow
\Omega ^1
\eqa
 Then it is natural to define the difference to be equal to:
\bqa
\mbox{Dim Ker }{\overline D}-\mbox{Dim CoKer }{\overline D}  =\left [
\begin{array}{c}
                             \mbox{ for (+) subalgebra } g-1 + \int F
\\
                             \mbox{ for (-) subalgebra } g-1 - \int F

                             \end{array} \right .
\eqa

Thus we have:
\bqa
\phi _0 ^{2(g-1) + \int F - \int F } = \phi _0 ^{2(g-1)}
\eqa
and the ratio of determinants is equal to:
\bqa
 \frac {det \left [2\phi ^{0} \mbox{{\bf I}}\right
]^{2(g-1)}_{(0)}}{det \left [2\phi ^{0} \mbox{{\bf I}}\right
]^{2(g-1)}_{(1)}} = \frac {1}{\phi ^0 }^{2(g-1)}
\eqa
The final result is given by the sum over all topologically different
classes of classical solutions. It is easy to see that in this case
we obtain the following expression:
\bqa
Z= \sum _m \int dy e^{2my-\epsilon y^2}\left (\frac {1}{y} \right
)^{2(g-1)} =\sum _n e^{\epsilon n^2} \left (\frac {1}{n} \right
)^{2(g-1)}
\eqa \label{SS1}
It is simple to show that all above calculations may be fulfilled  in
slightly more general case of the arbitrary potential for the field
$\phi $
\bqa
S = tr( \int  \phi F + \sum _k \epsilon _k \phi ^k )
\eqa
In this case we get the following natural generalization of
(\ref{SS1})
\bqa
S = \sum _m  e^{im\phi y + \sum _k \epsilon _k y^k } (\frac
{1}{y})^{2(g-1)} =\sum _n e^{\sum _k \epsilon _k n^k} (\frac
{1}{n})^{2(g-1)}
\eqa
\section{\bf GWZW theory}

Now we consider functional integral for two-dimensional theory with
the action of the form:
\bqa
S = k S_{WZW} (g) + \frac {ik}{2\pi } \int ( A_z g^{-1}\apr g + g\pr
g^{-1} A_{\z } +gA_z g^{-1} A_{\z } - A_z A_{\z} )
\eqa \label{GWZW}
Notice that these action is invariant under transformation:
\bqa
g \rightarrow hgh^{-1} \\
A_{\z} \rightarrow hA_{\z}h^{-1} + h\apr h^{-1} \\
A_{z} \rightarrow hA_{\z}h^{-1} + h\pr h^{-1}
\eqa
It will be useful to deal with slightly deformed theory:
\bqa
S = S_{\G} + \epsilon Tr \int d^2z (g-1)
\eqa
 which preserve all the symmetries of the action and functional
integral with this action may be considered as generating function of
the correlators in \G .

This theory is connected with YM theory considered in the previous
section. Introducing  the parametrization:
\bqa
g = e^{\frac {i\phi }{k}} \\
\epsilon = k^2 {\widetilde \epsilon }
\eqa
we get YM theory from \G when $k$ goes to $\infty $:
\bqa
S_{k \rightarrow \infty } = tr \int \phi F  +{\widetilde \epsilon }tr
\int \phi ^2
\eqa
It appears that if we write the measure for $A$ as an integral over
fermions $\Psi _{z},\Psi _{\z}$ similar to the case of YM theory \G
theory will exhibit BRST-like symmetry with fermionic parameter
$\lambda $:
\bqa
&& \delta A_{\mu} = \lambda \Psi _{\mu} \\
&& \delta \Psi _{\z} = \lambda (A^g _{\z} -A_{\z} ) \\
&& \delta \Psi _{\z} = - \lambda (A^{g^{-1}} _{z} -A_{z} ) \\
&& \delta g = 0
\eqa
To calculate exactly functional integral in \G we may deform the
action in the following way:
\bqa
&& \delta S = \left [ Q,V \right ] \\
&& V = \beta \int \Psi (A^g _{\z} - A_{\z}) +{\overline \Psi
}(A^{g^{-1}} _{z} - A_{z}) + \Psi \NA _{A,{\overline A^g }}F(A,
{\overline A^g })\\
&& \delta S = \beta \int |A^{g^{-1}} _{z} - A_{z}|^2 +|A^g _{\z}
-A_{\z}|^2  + (A^g _{\z} -A_{\z}) \NA F + \\
&& (A^{g^{-1}} _{z} - A_{z}) {\overline \NA } f \Psi (g^{-1}
{\overline \Psi }g - {\overline \Psi }) +  \\
&& {\overline \Psi }(g\Psi g^{-1} - \Psi ) + \Psi \delta ({\overline
\NA }F) + {\overline \Psi } \delta (\NA F)
\eqa
or just look at the zero locus of BRST transformations. In any case
we derive that this functional integral may be localized on the
solutions of of equations of motions with the exception of zero mode
of the field $g$ which should be treated exactly.
Consider equation of motions for \G:
\bqa
g^{-1}\apr g + g^{-1}A_{\z}g -A_{\z} =0 \\
g\apr g^{-1} + gA_{\z}g^{-1} -A_{\z} =0 \\
F_{z,\z }(A_z , g^{-1}\apr g + g^{-1}A_{\z}g ) =0
\eqa
We will restrict ourselves to the case of $SU(2)$. From the condition
for $g$ to be covariantly constant by usual arguments we get that we
may reduce $g$ to be constant matrix in $\sigma _3 $ subgroup of
$SU(2)$.

We  parametrize  the element of the group as;
\bqa
g = e^{i\phi \sigma _3 } = cos \phi + i \sigma _3 sin \phi
\eqa
and the classical action will have the form:
\bqa
S= k \int F \phi + \epsilon  cos \phi
\eqa
Taking into account the topological restrictions on total curvature
on the surface we have for particular topological sector :
\bqa
S_k ^{(m)} = 2\pi i km \phi + \epsilon cos\phi
\eqa
The second variation of the action is given by the expression:
\bqa
&& \delta ^2 S = k\int (\delta A_z g^{-1} \delta A_{\z} g - \delta
A_z \delta A_{\z})+ \int \NA _{A}(g^{-1} \delta g ) {\overline
\NA_{A^g} }(g^{-1}\delta g) \\
&& + \int \delta A {\overline \NA_{A^g} }(g^{-1}\delta g) + \int
g^{-1} \delta A_{\z} g \NA _{A}(g^{-1} \delta g ) +tr g g^{-1} \delta
g  g^{-1} \delta g
\eqa
It is useful to parametrize $g$ in terms of gauge transformations $h
= e^{\omega } = 1+\omega + \ldots $:
\bqa
&& \delta g = \delta (h^{-1} e^{\phi \sigma _3 }h) =\delta \phi
e^{\phi \sigma _3 }+e^{\phi}\omega -\omega e^{\phi} \\
&& (g^{-1}\delta g )^{+} = (1-e^{2\phi})\omega ^{+} \\
&& (g^{-1}\delta g )^{-} = (1-e^{-2\phi})\omega ^{-} \\
&& (g^{-1}\delta g )^{0} = \delta \phi ^0
\eqa
In this parametrization  the second variation of the action will be :
\bqa
&& \pr \delta \phi ^0 \apr \delta \phi ^0 +2\delta {\overline A^0}
\pr \delta ^0 +2\delta A^0 \apr \phi ^0 + \\
&& \left (e^{2\phi }-1\right )(\delta {\overline A^{+}}
+{\overline \NA}\omega ^{+} )(\delta {\overline A^{-}}
+{\overline \NA}\omega ^{-} ) +  \left (e^{-2\phi }-1\right )(\delta
{\overline A^{-}} \\
&& +{\overline \NA}\omega ^{-} )(\delta {\overline A^{+}}
+{\overline \NA}\omega ^{+} )
\eqa
The difference with YM case is mainly in one-loop contribution which
for \G has the following form (we take into account ghost
contribution ) :
\bqa
&& \nn \\
&& \frac {det _{0}(e^{2\phi } - 1 )(e^{-2\phi }-1){\bf
I}}{det_{1^{+}}((e^{2\phi } - 1 ){\bf I})det_{1^{-}} ((e^{-2\phi
}-1){\bf I})}  \\ && \nn
\eqa
where $det_0 ({\bf I}) $ is determinant of the unit operator on the
space of 0-forms,and $det_{1^{\pm}} ({\bf I}) $ is the determinant on
the space of one-forms proportional to $ \sigma ^{\pm} $. Using the
same arguments  as in the case of YM theory we have for the ratio of
the determinants:
\bqa
&& (e^{2\phi } - 1 )^{g-1 +\int F }(e^{-2\phi }-1)^{g-1 -\int F } =
\\
&& = \frac {1}{(2sin \phi )^{2(g-1)}} e^{c_v \int F \phi }
\eqa
where dual Coxeter number is $c_v =2$ for $SU(2)$.

Thus we get the well known renormalisation:
\bqa
k \rightarrow k + c_v
\eqa
The final step of calculation looks as following:
\bqa
Z(\epsilon ) = \sum _m \int e^{(k + 2)m\phi +\epsilon (cos \phi
-1)}\frac {2^{g-1} (k+2)^{g-1}}{(2sin \phi )^{2(g-1)}} d\phi = \\
= \sum ^{k+1}_{l=1}e^{\epsilon (cos \frac {2\pi l}{k+2} -1)}\frac
{2^{g-1} (k+2)^{g-1}}{(sin \frac {2\pi l}{k+2}  )^{2(g-1)}}
\eqa
where $ (k+2)^{g-1}$ is the volume of the Jacobian divided on
normalized volume of zero mode and $2^{g-1}$ is due to center of the
group. This is exactly Verlinde formula for dimension of the space of
conformal blocks.

It is necessary to comment on the form of the action (\ref{GWZW}). We
have considered topologically nontrivial bundles over Riemann
surface. In this case there is no well defined connection and we need
covariant generalization of (\ref{GWZW}) for this case. It is simple
to verify that the following action have all necessary properties.

\bqa
&& S_{\G} = \int _{\Sigma} g^{-1} \NA _A g  g^{-1} \NA _A +  \\
&& \int _{B} (g^{-1} \NA _{A} g  g^{-1} \NA _{A} g g^{-1} \NA _{A} g
+ F(A)(g \NA _{A} g^{-1} - g^{-1} \NA _{A} g) \nn
\eqa
where $\NA _A $ is the covariant derivative and $B$ is the three
dimensional manifold which has as the boundary two dimensional
surface $\Sigma $.
\section{\bf Conclusions}
We have shown how to get exact answers for correlation functions in
\G model for compact group. It would be interesting to compare these
computations with the possible generalization of Migdal approach
\cite{Mig} to \G .It seems that required generalization may be
obtained more or less by using quantum finite dimensional groups
instead of classical finite dimensional groups. The form of the
Verlinde formula seems support this suggestion.

In other direction it would be nice to generalize the calculations
described above to the case of non-compact groups connected with
theories of gravitational type. Another appealing possibility is
two-loop groups closely related with integrable systems \cite{G}.
\vskip 1cm
{\bf \noindent Acknowledgements } \\
\noindent
I am grateful to A.Niemi for hospitality in Uppsala.

\end{document}